%% file: final.tex
\documentclass{article}

\input{preamble}

\title{Quantifying SMT Solver Instability for Formal Verification\footnote{\protect\input{copyright}}
}

\author{
    Daniel Evans\affmark[1], Rylie Hansen\affmark[1], and John Shorter\affmark[1]\\
    \affmark[1]Computer Science Department\\
    St. Olaf College\\
    Northfield, MN\\
    \email{\{evans11,hansen42,shorte1\}@stolaf.edu}
}

\begin{document}
\maketitle

\begin{abstract}
Program verification has become commonplace in software development projects, and much of its feasibility can be attributed to the potential of SMT solvers to automate model checking and other verification techniques. However, SMT instability, a phenomenon where semantically equivalent permutations of source code take much longer to solve or timeout, has been found to interrupt workflows and make software engineers more likely to avoid verification. Previous efforts to measure SMT instability have had difficulty quantifying it due to the abnormal runtime distributions it produces. In this paper, we use the coefficient of variation to model instability across an extensive dataset of SMT queries from three software domains. Our results show that SMT instability is significant in each area, and that it increases for queries that take longer on average. However, using a novel method to quantify instability, we show that a different approach is possible: if a solution were to optimize the structure of queries for runtime, rather than variance, instability could be mitigated while also improving overall solver performance. 
\end{abstract}

\section{Introduction}
Satisfiable Modulo Theories (SMT) is a class of problems that seek to determine whether or not a solution to a formula exists. As an expressive method for representing formal logic problems, they are useful for software engineering tools such as model checkers, symbolic executors, and proof-oriented programming languages, which structure facts about software as SMT queries and solve them using programs called SMT solvers \cite{luckcuck_formal_2019,lattuada_verus_2023}. Solving an SMT query is an NP-complete problem, and the solving process can be a time consuming step for formal verification tools; thus, many solvers use heuristics to navigate the solution space \cite{kroening_decision_2008}.

Heuristics make the problem more feasible, but they also introduce a phenomenon referred to as the butterfly effect, proof brittleness, or proof instability \cite{bordis_free_2025}. Small changes to source code or logically equivalent permutations to an SMT query can lead to significantly different runtimes for the solver \cite{zhou_mariposa_2023}. This is a widely recognized problem in academia and large-scale software production because any small modification to a script---for example, editing a naming convention---has the potential to make verification intractable \cite{zhou_cazamariposas_2025}.

While the existence of proof instability has been well documented, the full scope of the problem is not well known, and most methods of mitigating it are either ad-hoc or incomplete \cite{zhou_context_2024}. In this work, we use the coefficient of variation (CV) and other statistical methods to quantify instability within an extensive benchmark study of 60,000 SMT query forms across three software domains. We find that instability is significant for complete samples, and our results provide strong evidence that instability continues to be a problem for distributions with timeouts. Finally, we demonstrate that intelligently mitigating instability---while optimizing for runtime---can reduce average solver runtime, a step toward addressing instability.\\

\section{Background}
The phenomenon of instability (also called proof fragility or proof brittleness) among SMT queries has been observed as a challenge in many large-scale, industry verification-based projects \cite{dodds_formally_2022}, including those from Microsoft \cite{hawblitzel_ironclad_2014,lattuada_verus_2024} and Amazon \cite{Chakarov2025}. 

In order to formally investigate the prevalence of instability in proofs generated by Dafne, a verification programming language, \cite{zhou_mariposa_2023} developed the Mariposa tool and benchmark. They introduce the conception of a query as one member of a distribution of equivalent queries that have been mutated in ways that preserve semantic meaning, which we build off of in our work. Statistical approaches to distributions of runtimes are made challenging by the reality that the underlying distribution is not known---i.e. it is not normal and often has very long tails---and frequently cannot be estimated due to timeout constraints. Success rate avoids some of these challenges by treating solver results as binary, rather than a continuous measure of time, but in doing so loses the ability to quantify or compare the instability of one query set with another. 

This work was continued in Shake \cite{zhou_context_2024} and Cazamariposas \cite{zhou_cazamariposas_2025}, the former a method that uses the empirically identified technique of context pruning to reduce instability, and the latter a tool that identifies candidate qualifiers for instability and suggests repairs. An important observation from these works is the contribution of problematic axioms to instability, which is taken advantage of by \cite{dodds_formally_2022}. 

\cite{amrollahi_towards_nodate} determined that, if a normalized or canonical form could be found for each logically equivalent query, it would eliminate instability when making irrelevant changes. While finding such a form is an NP-complete task in itself, they produce a normalizing algorithm that works well in most cases. They represent runtimes as a distribution of penalized runtime (PR-2) scores, where a query-solver pair is evaluated by runtime where the solver succeeds, and twice the timeout time where it fails. Then, Median Absolute
Deviation (MAD) in PR-2 scores is used to measure instability, which, unlike success rate, allows for quantification and comparison. However, it is impossible to say how much the true runtime would have deviated from the PR-2 score for queries that timeout, so the statistical significance is unknown. Moreover, MAD does not quantify instability relative to the difficulty of a query set, meaning that one query set may appear to be more unstable than another simply because it takes longer to compute, on average.

\section{Methodology}

\subsection{Quantifying Instability}
As in \cite{zhou_mariposa_2023}, we use the quality of \textit{semantic equivalence}. Given SMT queries $Q$ and $Q'$, we say that $Q$ is semantically equivalent to $Q'$ if, for any assignment $\phi$, $Q(\phi)\iff Q'(\phi)$. \cite{amrollahi_towards_nodate} identified several ways to mutate a query and preserve semantics, including renaming variables, shuffling the clauses of commutative operators, like conjunction or addition, and swapping symmetric operators, like less than and greater than. Given some query $Q$ and a semantic-preserving mutation $M$, we define the \textit{mutation set} of $Q$ to be the set of all semantically equivalent queries that can be derived by applying $M$ to $Q$. 

By treating the mutant set of $Q$ as a population to be sampled from, statistical analysis of the variance in runtimes that result from applying some solver $S$ to members of a query's mutant set can describe proof instability. Although instability in the field may involve non-equivalent queries, showing it for equivalent ones suffices.

Variance among a population of runtimes is often estimated using the standard deviation statistic, denoted $\sigma$, of a sample, and previous work has used standard deviation as a proxy for instability \cite{zhou_mariposa_2023}. However, this makes queries whose mutants typically take longer appear more unstable, when the difficulty of a query and its instability should be considered independently. We use the mean runtime, denoted $\mu$, as a proxy for the difficulty of a query's mutant set. The coefficient of variation (CV) statistic, defined as $$CV=\frac{\sigma}{\mu},$$
accounts for this by evaluating the variance of runtimes relative to the difficulty of the mutant set \cite{ross2020introduction}. 

Constructing a confidence interval would indicate the trustworthiness of the estimated CV, but the underlying distribution of runtimes is not consistent. To overcome this, we use a nonparametric bootstrapping approach to estimate the variability of the CV, as described in \cite{puth_variety_2015}. Randomly selecting from a sample of mutant query runtimes with replacement produces what is called a bootstrap, from which we calculate a resampled CV. By generating bootstraps many times, we get a distribution of CVs and an estimation of variance, from which we construct the desired confidence interval. 

In analyzing our results, we say that a query is unstable if the lower bound of the 95\% confidence interval around its CV is greater than or equal to one. That is, a query is unstable when its mutant set has a standard deviation greater than or equal to its mean, with 95\% confidence. The choice of threshold is arbitrary, but the ability to choose any threshold demonstrates the power of this method, as it can be tailored to any desired amount in practice.

\begin{figure}[htbp]
    \centering
    \includegraphics[width=\textwidth]{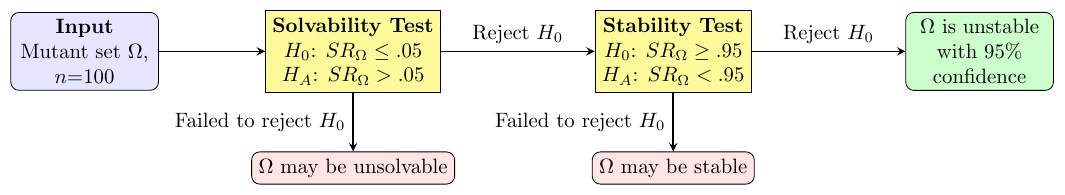}
    \caption{\textbf{Categorization of Unstable When $SR<1$}}
    \label{fig:unstable_flow}
\end{figure}

Because it is necessary to impose a time limit on solvers, some samples have missing runtime data. It is not possible to make good statistical estimations about the variance of these distributions, so, in these cases, we resort to a simplified version of the method used by \cite{zhou_mariposa_2023}. That is, we define the success rate (SR) of a sample to be the proportion of mutant queries for which the solver does not timeout. Then, if a solver runs out of time on less than 5\% of the queries, the original query is considered \textit{unsolvable} and not unstable, while an SR greater than .95 is considered \textit{stable} and not unstable. In order to be statistically confident in concluding that an original query is unstable, we perform a z-test of proportions for unsolvability and stability. This process is summarized in Figure \ref{fig:unstable_flow}.

\subsection{Study Design}
\subsubsection{Domain}
We limit the scope of our analysis to the cvc5 \cite{barbosa_cvc5_2022} and Z3 \cite{de_moura_z3_2008} SMT solvers. Both solvers are widely used in industry and academia, and previous work on instability has focused on them. The SMT queries we benchmarked were pulled from three different sources: (i) bounded model checking, (ii) symbolic execution, and (iii) competition. 

Bounded model checkers (BMCs) build finite state models of programs and prove that they cannot reach states encoded as error states. They frequently build these models using SMT queries \cite{beyer_software_2022}. Symbolic execution (SymEx) engines use SMT queries to represent the end behavior of programs. Solving these queries proves that a particular result---e.g. an error---is possible or impossible. All of the BMC and SymEx queries used in this study were collected by \cite{leeson_sibyl_2023} in a public dataset, using the ESBMC model checker and KLEE symbolic execution engine on the GNU coreutils and other real world programs. In addition to these two areas, we used SMT queries from the 2025 International Satisfiability Modulo Theories Competition (SMT-Comp), a yearly event that has developers test SMT solvers on hundreds of thousands of competitions. Previous work has already demonstrated the existence of instability in SMT-Comp queries \cite{zhou_mariposa_2023}, but BMC and SymEx have not been explored directly. 

\subsubsection{Query Selection}

For BMC, SymEx, and SMT-Comp, we gathered two subsets of queries. The process for BMC and SymEx differed from that of SMT-Comp because we included queries from specific SMT-Comp logics.

The first selection consisted of a random sample. We randomly chose 100 queries from both BMC and SymEx and 150 queries from SMT-Comp, equally sourced across 15 logics. Every query was run using both solvers, and, if an error occurred, it was replaced with a new randomly chosen query. SMT-Comp has different subsets of query logics to fulfill, while BMC and SymEx contain no such divisions.

The second of our two selections consisted of difficult queries, which we defined as queries that timed out for cvc5, Z3, or both. We selected 100 hard queries from BMC and SymEx and 120 hard queries from 6 logics in SMT-Comp. For SMT-Comp, the remaining queries were drawn from those timing out on either solver. For BMC and SymEx, they were split evenly between CVC5-only and Z3-only timeouts, with reallocation if needed.

\subsubsection{Query Shuffling}
For the purposes of this study, we modified the official SMT Query scrambler used in the SMT-Comp by limiting its scope and implementing a basic normalization process. The existing scrambler applies many different semantic-preserving mutations, so we modified this behavior to only shuffle assertions, thus limiting the scope of the data. Other works have analyzed instability caused by other mutations, like shuffling the clauses of commutative functions \cite{amrollahi_towards_nodate}. For the sake of consistency, we normalized variable names, which prevents the effects of declarations from affecting instability.

\section{Results}
\subsection{Prevalence of Instability}
Table \ref{table:nonlin} contains the results of each solver, domain, and selection category. For every original query in the SymEx hard category, CVC5 had an SR less than one, so an accurate CV could not be calculated. In most categories, the proportion of unstable queries is small, which is consistent with previous results \cite{zhou_mariposa_2023}. However, even a small percentage of inconsistent queries can cause enough failure to limit the use of formal verification by software developers \cite{zhou_cazamariposas_2025}, so these results are significant, particularly in categories with more than 10\% unstable queries like Z3 solving hard BMC queries.

\begin{table}[ht]
\caption{SMT Solver Benchmark Results}
\label{table:nonlin}
\centering
\resizebox{\textwidth}{!}{%
\begin{tabular}{lllrrlrr}
\toprule
Domain & Solver & Selection & Mean & $CV$ & $CV$ 95\% CI & $SR$ & Unstable Proportion \\
\midrule
BMC & cvc5 & hard-scrambled & 1114 & 0.06 & [0.04,0.07] & 0.09 & 0.09 \\
 &  & pure-random-scrambled & 232 & 0.15 & [0.07,0.18] & 0.83 & 0.06 \\
 & Z3 & hard-scrambled & 751 & 0.61 & [0.38,0.72] & 0.43 & 0.29 \\
 &  & pure-random-scrambled & 72 & 0.22 & [0.15,0.26] & 0.95 & 0.08 \\
SMT-Comp & cvc5 & hard-scrambled & 1133 & 0.24 & [0.14,0.29] & 0.08 & 0.12 \\
 &  & pure-random-scrambled & 250 & 0.12 & [0.08,0.15] & 0.80 & 0.06 \\
 & Z3 & hard-scrambled & 1111 & 0.04 & [0.02,0.05] & 0.11 & 0.14 \\
 &  & pure-random-scrambled & 182 & 0.20 & [0.1,0.24] & 0.87 & 0.05 \\
SymEx & cvc5 & hard-scrambled & 1101 & NA & [NA,NA] & 0.12 & 0.07 \\
 &  & pure-random-scrambled & 157 & 0.13 & [0.1,0.15] & 0.96 & 0.09 \\
 & Z3 & hard-scrambled & 11 & 0.19 & [0.16,0.22] & 0.99 & 0.01 \\
 &  & pure-random-scrambled & 1 & 0.18 & [0.15,0.21] & 1.00 & 0.00 \\
\bottomrule
\end{tabular}
}
\end{table}

Moreover, comparing CV, our contribution, to SR, the previously used method for measuring instability, provides various insights. In many cases, a low SR is associated with high average runtimes and vice versa. This shows how reliant SR is on the timeout threshold; easier queries may appear stable at twenty minutes, while having a high amount of variance that would lead to them being considered unstable with some other time limit. CV gives a more accurate indication of instability than the proportion of queries marked as unstable for a given category. For example, among randomly selected BMC queries, the average success rate of Z3 was about 95\%, which makes the unstable proportion appear relatively low at 0.08. The CV, however, shows that these queries still varied by about 22\% on average, which could be very significant under a lower time limit. 

\subsection{Relationship Between Query Difficulty and Instability}
Another advantage of being able to quantify instability is being able to investigate its relationship to other variables. We applied linear regression to all mutant sets with a success rate of one, and we found that the logarithm of CV is a very significant positive predictor of the logarithm of average runtime (${\rm t=10.5,df=263,p<.001}$), which provides strong evidence that, within the domain of our analysis, more difficult queries tend to be more unstable. This is further supported by the fact that CVs tended to be higher in the hard selections as opposed to the randomly selected. Furthermore, difficult queries for which the CV could not be estimated are likely to exhibit significant variation, on average. 

\begin{figure}[htbp]
    \centering
    \includegraphics[width=1.\textwidth]{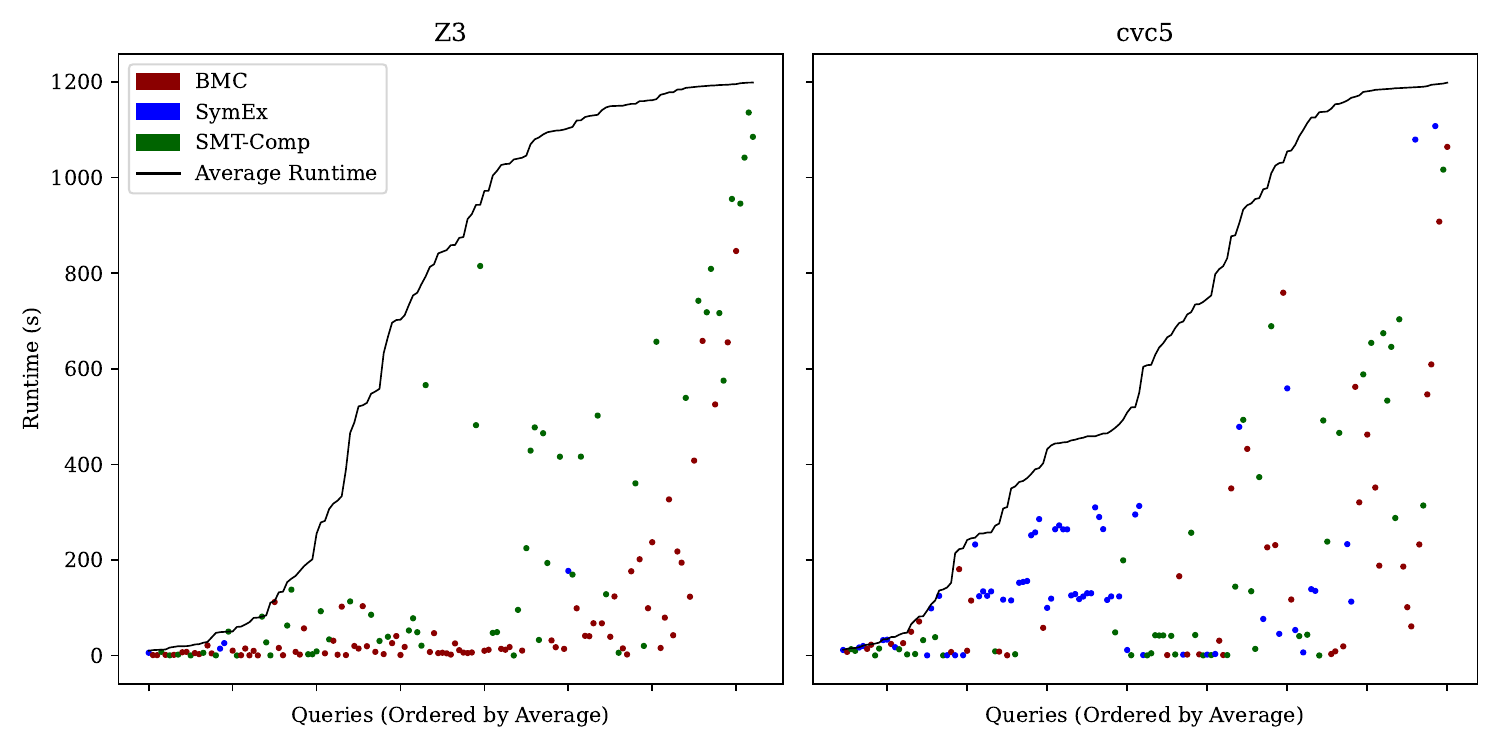}
    \caption{\textbf{Difference Between Mean and Minimum Runtimes}: Mutant sets with average times of less than 10 or equal to 1200 seconds are excluded. Z3 in particular exhibits high potential for time savings.} 
    \label{fig:separation}
\end{figure}

\subsection{Leveraging Instability}
When a software engineering tool generates an SMT query, that query is not a random permutation within a broad set of semantically equivalent queries; instead, the query is produced according to some internal algorithm. Naturally, the concern of this algorithm is in generating rigorous proofs that verify qualities of software, not in creating a particular structure that is well-suited for the heuristics of SMT solvers. Thus, if the developer hasn't considered this, the query the tool produces may as well be truly random, and has previously been treated as such.

The prevalence of instability, particularly among difficult queries, suggests that queries that are structured more intelligently could save time when they reach the SMT solver. This is visualized in Figure \ref{fig:separation}, where minimum runtimes within a query's mutant set are plotted beneath the original runtime. In some cases, a well-structured query's runtime is negligible, while the original query takes nearly sixteen minutes to solve.

Quantifying instability using CV allows us to be more exact. For example, we can say that, with a confidence level of 0.95, restructuring a hard BMC query could decrease Z3 runtimes by as much as 38\% on average. Previous work suggests that the runtime of the original query typically correlates with the center of its mutant set \cite{zhou_mariposa_2023}, which means that the average query runtime is a good stand-in for the original runtime. Our results show that future work should consider mitigating instability by optimizing query structure for runtime, rather than just success rate or variance, as this would have a stabilizing effect with the added benefit of decreased runtimes. 

\section{Conclusion}
We investigated SMT instability in three relevant software development areas, two of which had previously been unexplored. In addition to using the previous metric of success rate, we applied the coefficient of variation to the runtimes of 60,000 mutant queries, and our results show that instability is an issue for both the cvc5 and Z3 solvers in every domain. By fitting a linear model to CV, we demonstrate that it is possible to predict variation in queries and that the difficulty of a query is a significant positive predictor of instability. Thus, it is reasonable to expect that, for difficult queries where timeouts occur, we would find significant instability as well.

Finally, our results show that query instability may be an opportunity as much as it is a challenge. If an efficient method could be developed for restructuring queries in a way that is better suited to the internal heuristics of solvers, it could mitigate instability while also offering time savings. While work is currently being done on developing bespoke methods identified in the field for mitigating instability \cite{zhou_cazamariposas_2025}, we conjecture that optimizing for runtime would be a task well-suited for machine learning models that are capable of reasoning about the complex structure of algorithms and program data. 

\section{Acknowledgments}
The authors would like to express gratitude to Dr. Will Leeson, Professor of St. Olaf College, and Nicholas Phair, PhD Student at the University of Virginia, for supporting this research.
\medskip

\printbibliography

\end{document}

%% file: preamble.tex
\usepackage[
  paperheight=8.5in,
  paperwidth=5.5in,
  left=10mm,
  right=10mm,
  top=20mm,
  bottom=20mm]{geometry}
\usepackage[utf8]{inputenc}

\usepackage{graphicx}
\usepackage{wrapfig}
\usepackage[bottom]{footmisc}
\usepackage{listings}
\usepackage{enumitem}

\usepackage{wrapfig}
\usepackage{ragged2e}

\usepackage{array}
\usepackage[table]{xcolor}
\usepackage{multirow}
\usepackage{booktabs}
\usepackage{hhline}
\definecolor{palegreen}{rgb}{0.6,0.98,0.6}

\usepackage{amsmath}
\usepackage{amssymb}
\usepackage{multicol}
\usepackage{lipsum}
\usepackage{hyphenat}
\PassOptionsToPackage{hyphens}{url}
\usepackage{url}

\usepackage{rotating}

\usepackage[T1]{fontenc}
\usepackage{textcomp}
\usepackage{listings}
\usepackage{setspace}

\newcommand*{\affmark}[1][*]{\textsuperscript{#1}}
\newcommand*{\email}[1]{\small{\texttt{#1}}}

\renewcommand{\footnoterule}{%
  \kern -3pt
  \hrule width \textwidth height 0.5pt
  \kern 2pt
}

\date{}

\usepackage{titlesec}
\titleformat*{\section}{\large\bfseries}
\titleformat*{\subsection}{\normalsize\bfseries}
\titleformat*{\subsubsection}{\normalsize\bfseries}

\usepackage[backend=biber,style=ieee,sorting=none,sortcites=true]{biblatex}

\usepackage[ruled,vlined]{algorithm2e}
\usepackage{soul}

%% file: copyright.tex
Copyright \copyright 2022 by the Consortium for Computing Sciences in Colleges.
Permission to copy without fee all or part of this material is granted provided
that the copies are not made or distributed for direct commercial advantage,
the CCSC copyright notice and the title of the publication and its date appear,
and notice is given that copying is by permission of the Consortium for
Computing Sciences in Colleges.  To copy otherwise, or to republish, requires
a fee and/or specific permission.